\documentclass{hep99}
\usepackage{epsfig}
\def\zero{{\scriptscriptstyle 0}}

\def\Z0{\ensuremath{Z^\zero}}
\def\SU2U1{{\rm SU}(2)\times{\rm U}(1)}

\mathchardef\qsm=63
\mathchardef\pls=43
\mathchardef\mns=512
\mathchardef\plm=518
\mathchardef\eql=61
\mathchardef\smallleft=300
\mathchardef\smallright=301
\mathchardef\perslsh=47
\mathchardef\les=316
\mathchardef\gre=318
\mathchardef\leq=532
\mathchardef\grq=533
\chardef\usc=95
\chardef\til=126
\def\sqr#1#2#3{{\vcenter{\hrule height.#3ex\hbox{\vrule width.#2ex height#1ex
    \kern#1ex\vrule width.#3ex}\hrule height.#2ex}}}

\def\angleto{\vrule width.035em height2.1ex depth-.56ex\unskip\kern-.6ex\to}
\def\perchc#1{{\raise.4ex\hbox{$\mkern4mu#1{\it\perslsh}_
             {\mkern-5mu\scriptscriptstyle{{\rm o}\!{\rm o}}}^
             {\mkern-12.8mu\scriptscriptstyle{\rm o}}$}}}

\catcode`\@=11 % @ signs are now treated as letters
\def\parenbar{\mathpalette\p@renb@r}
\def\p@renb@r#1#2{\vbox{%
  \ifx#1\scriptscriptstyle \dimen@.7em\dimen@ii.2em\else
  \ifx#1\scriptstyle \dimen@.8em\dimen@ii.25em\else
  \dimen@1em\dimen@ii.4em\fi\fi \offinterlineskip
  \ialign{\hfill##\hfill\cr
    \vbox{\hrule width\dimen@ii}\cr
    \noalign{\vskip-.3ex}%
    \hbox to\dimen@{$\mathchar300\hfil\mathchar301$}\cr
    \noalign{\vskip-.3ex}%
    $#1#2$\cr}}}
\catcode`\@=12 % @ signs are no longer letters

\newbox\struttbox
\setbox\struttbox=\hbox{\vrule height1.65ex depth.485ex width0pt}
\def\strutt{\relax\ifmmode\copy\struttbox\else\unhcopy\struttbox\fi}
\def\stru#1#2{\relax\ifmmode\hbox{\vrule height#1 depth#2 width0pt}
\else\vrule height#1 depth#2 width0pt\fi}
\def\ronum#1{\uppercase\expandafter{\romannumeral#1}}
\def\ronuml#1{\expandafter{\romannumeral#1}}
\DeclareMathAlphabet{\mathbf}{OT1}{cmr}{bx}{sl}

\newcommand{\CPC}[3]{Comp. Phys.\ Commun.\ {\bf#1} (#2) #3}
\newcommand{\PR}[3]{Phys.\ Rev.\ {\bf#1} (#2) #3}
\newcommand{\ZP}[3]{Z.\ Phys.\ {\bf#1} (#2) #3}
\newcommand{\NP}[3]{Nucl.\ Phys.\ {\bf#1} (#2) #3}
\newcommand{\PL}[3]{Phys.\ Lett.\ {\bf#1} (#2) #3}
\begin{document}

\title{Prompt Photon Production \\ and Observation of \\
Deeply Virtual Compton Scattering}

\author{P R B Saull$^1$ \\
         {\em for the ZEUS Collaboration}}

\address{$^1$ Deutsches Elektronen-Synchrotron DESY, Notkestrasse 85, 22603 Hamburg, Germany\\
E-mail: {\tt patrick.saull@desy.de}}

\abstract{Two recent results in $ep$ physics using the ZEUS detector at HERA 
are discussed: a measurement of the prompt photon production cross section in 
photoproduction, and the first observation of Deeply Virtual Compton Scattering
in DIS.}

\maketitle

\section{Introduction}
Two types of real photon 
production processes in $e^+p$ collisions which are
currently studied using the ZEUS detector at HERA are 
prompt-$\gamma$ production in photoproduction ($\gamma p$) and Deeply Virtual 
Compton Scattering (DVCS) in deep inelastic scattering (DIS). 

\begin{figure}[h]
\epsfxsize=2.0cm
\centerline{\epsfbox{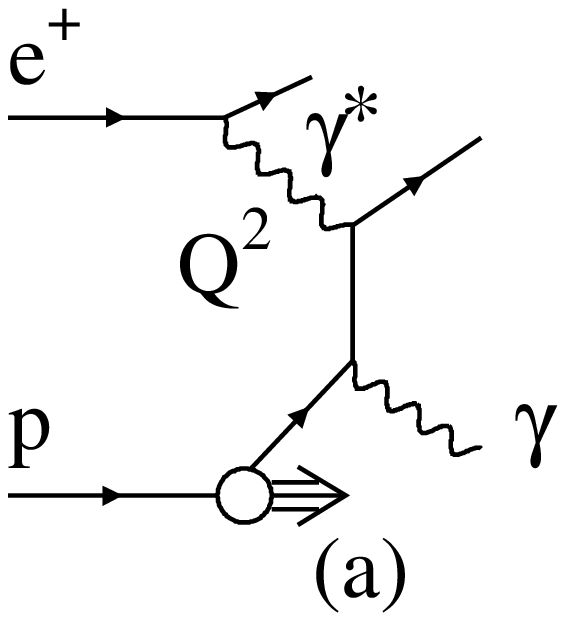}
\hspace{1.0cm}
\epsfxsize=2.0cm
\epsfbox{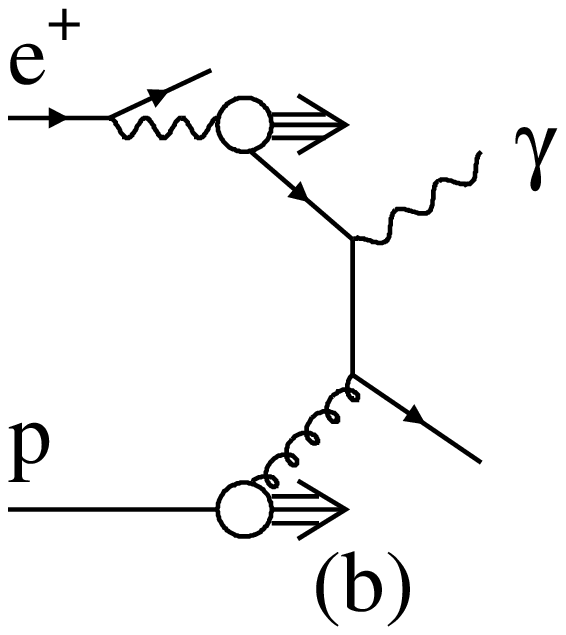}}
\caption{Examples of (a) direct and (b) resolved prompt-$\gamma$ production 
at leading order in $ep$ collisions.}
\label{fig_intro_diagrams_pp}
\end{figure}

Prompt-$\gamma$ production in 
$\gamma p$ interactions is the production of 
a real $\gamma$ directly from the hard interaction of 
a quasi-real $\gamma$ (invariant mass 
$Q^2 \approx 0$~\footnote{$Q^2 = 
-(e-e')^2$, where $e$ and $e'$ are the initial and final positron
four-momenta, respectively.})
with the proton. At leading order, 
two kinds of $\gamma p$ processes can be defined: direct, 
where the $\gamma$ participates entirely in the hard 
interaction, and resolved, where the $\gamma$ first fluctuates into 
a hadronic system and a parton from this system then enters into the 
hard interaction. 
Examples 
are depicted in Fig.~\ref{fig_intro_diagrams_pp}. Prompt-$\gamma$ production 
is less influenced by hadronisation effects 
than, e.g., dijet production, although
its production rate is down by a factor $\alpha / \alpha_S$  
in comparison. It
is interesting because of its sensitivity to the parton 
density of the $\gamma$. The
measurements can also be used to test next-to-leading order (NLO)
perturbative QCD (pQCD).

\begin{figure}[h]
\centerline{
%\hspace{1.0cm}
\epsfxsize=5.0cm
\epsfbox{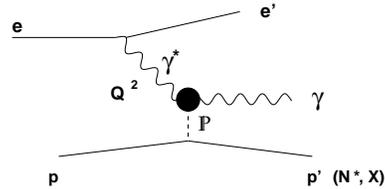}}
\caption{Diagram for DVCS.}
\label{fig_intro_diagrams_dv}
\end{figure}

DVCS~\cite{earlywork}, 
depicted in Fig.~\ref{fig_intro_diagrams_dv}, is the hard 
diffractive scattering of a $\gamma$ off a proton, 
$e^+p \rightarrow e^+ \gamma p$.
This is a process which has never been seen before at high energies, 
but which is predicted by FFS~\cite{ffs} to have a fairly high counting rate.
DVCS is an exciting process because of its potential for accessing the
skewed parton distributions (SPD's) of the proton. SPD's, which quantify
two-particle correlations in the proton, are a generalisation of 
the usual proton parton distributions to the case where the 
squared momentum transfer to the proton is non-zero. 
The advantage of DVCS 
over processes where a hadron is diffractively 
produced is two-fold: the theoretical uncertainty 
in the hadronic wave function is avoided, and the DVCS rate is 
predicted to be less suppressed 
by a factor of $Q^2$~\cite{freund}. Furthermore,
its final state is identical to that of QED Compton 
scattering (QEDC), and interference of the two processes
% (see Fig.~\ref{fig_intro_diagrams_bh}) 
potentially allows the measurement of the real part of a QCD amplitude.
%%
%\begin{figure}[h]
%\centerline{
%%\hspace{0.1cm}
%\epsfxsize=6.0cm
%\epsfbox{../dvcs/bethe_heitler.eps}}
%\caption{Interfering QEDC process.}
%\label{fig_intro_diagrams_bh}
%\end{figure}
%%

A common feature of these two studies at ZEUS is that each involves 
the detection of a $\gamma$ in the barrel part of the calorimeter (BCAL),
which has its electromagnetic (EM) section segmented into $5\times20$~cm$^2$ 
cells. A potential background contribution to
$\gamma$ candidates, i.e. those tagged 
EM clusters not associated with a track in the 
central tracking detector, arises from $\pi^0$ and $\eta$ production.
However, these particles tend to produce broader clusters in the BCAL
than  single $\gamma$'s. Two shower shape variables~\cite{shapevar}
are employed to exploit this difference: $Z_{width}$, which is the 
energy-weighted average of the width of the EM cluster in the 
$Z$-direction~\footnote{
The ZEUS coordinate system is defined as right-handed 
with the $Z$-axis pointing in the forward (proton beam) 
direction.
The origin is at the nominal 
$ep$ interaction point, and the polar angle $\theta$ 
is defined with respect to the positive $Z$-direction.} (the direction
in which the BCAL is most finely segmented); and $f_{max}$, which is 
the fraction of the EM cluster energy carried by the most energetic cell 
in the cluster.

\section{Prompt Photon Production}

\begin{figure}
\vspace{-0.5cm}
\centerline{
\epsfxsize=5cm
\epsfbox{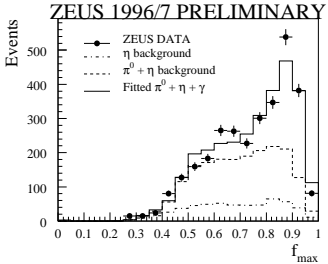}}
\vspace{-.5cm}
\caption{$f_{max}$ distribution for prompt-$\gamma$ candidates.}
\label{fig_fmax}
\end{figure}

\begin{figure}[h]
\centerline{
%\hspace{0.1cm}
\epsfxsize=3.5cm
\epsfbox{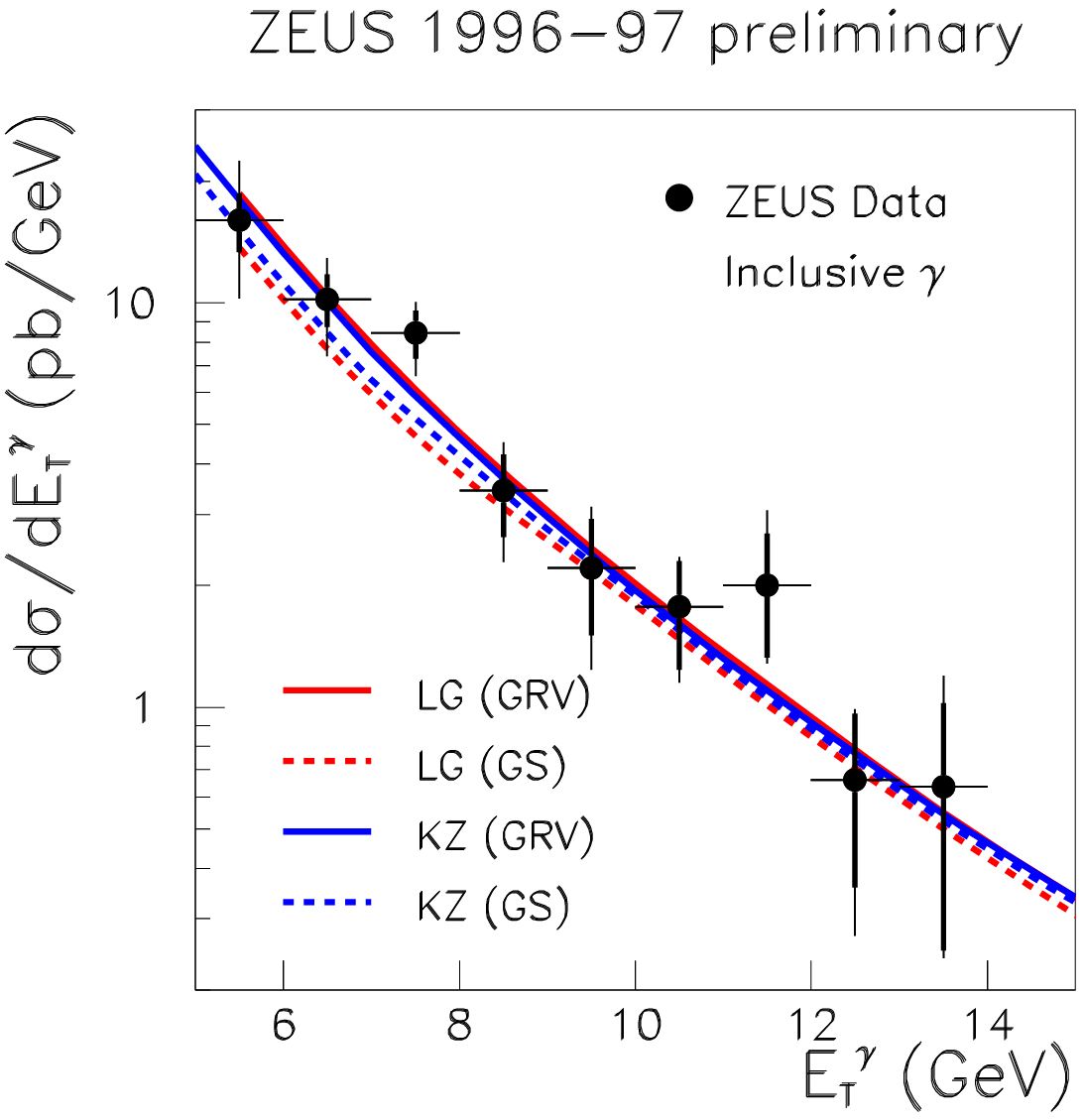}\hspace{.2cm}
\epsfxsize=3.5cm\epsfbox{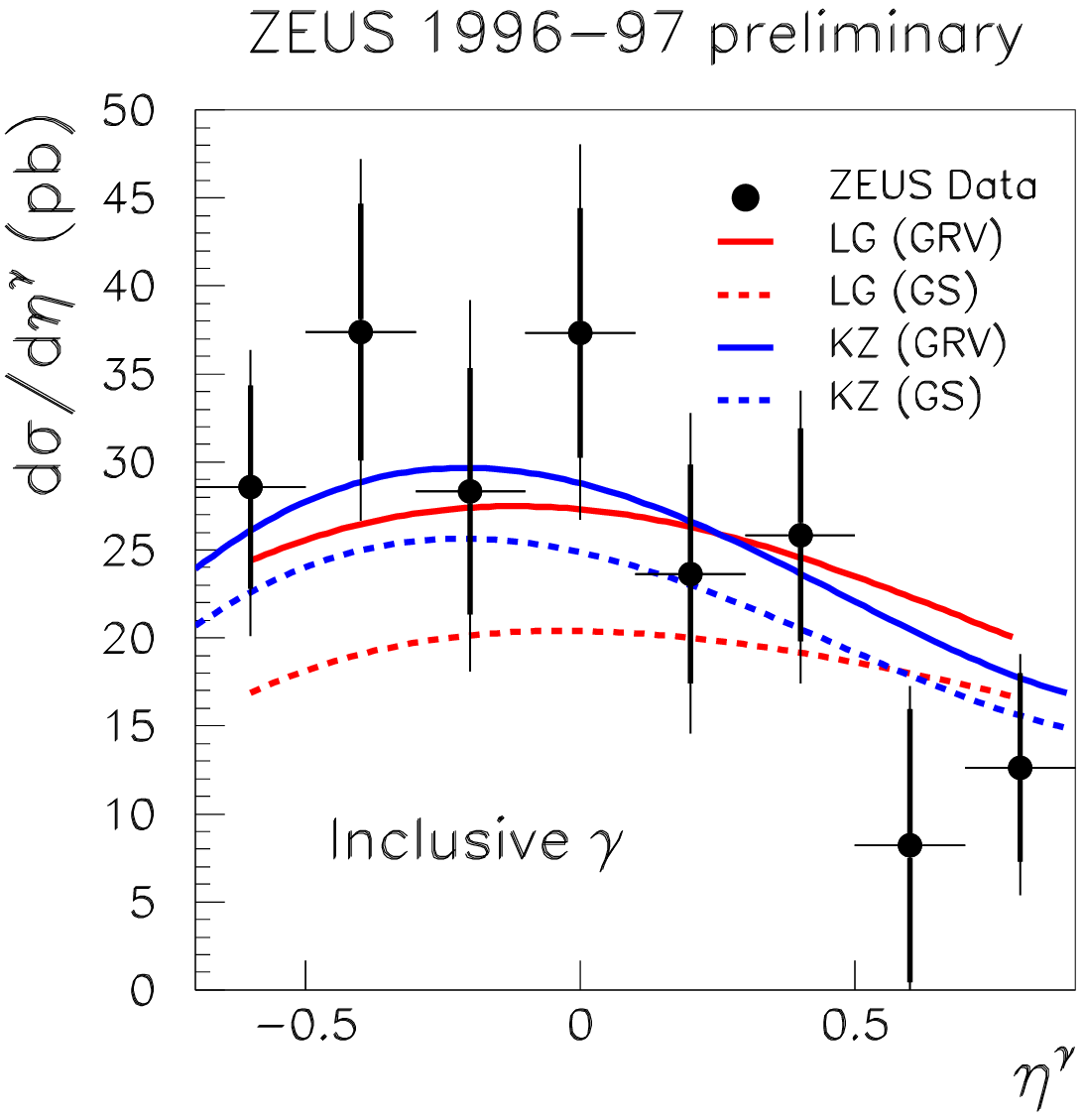}}
\caption{Prompt-$\gamma$ cross sections as a function of 
$E_T^{\gamma}$ and $\eta^{\gamma}$, with NLO predictions overlaid.}
\label{fig_pp_nlo}
\end{figure}

The $f_{max}$ distribution for the 1996/97 prompt-$\gamma$ photoproduction 
analysis~\cite{eps99papers} is shown in Fig.~\ref{fig_fmax}.
Overlaid are the expected contributions from $\eta$, $\pi^0$, and $\gamma$.
A $\gamma$ signal at high $f_{max}$ values is evident. 
The prompt-$\gamma$ production cross section for $\gamma$ transverse energies
$E_T^{\gamma}>5$~GeV and $\gamma$ rapidity interval
$-0.7 \le \eta^{\gamma} \le 0.9$ is shown in  Fig.~\ref{fig_pp_nlo} as a
function of $E_T^{\gamma}$ and $\eta^{\gamma}$. 
Overlaid are the NLO predictions of two groups of theorists, LG~\cite{lg} 
and KZ~\cite{kz}, each using two different $\gamma$ 
parton density parametrisations: GRV~\cite{grv} and GS~\cite{gs}.
The predictions are in reasonable agreement with the data, although those
based on the GS parametrisation tend to be low. This
demonstrates the sensitivity of this type of analysis to
the $\gamma$ parton density. With more data and further theoretical progress
the prompt-$\gamma$ process will provide a valuable tool for studying NLO pQCD 
and measuring the $\gamma$ parton density.

\section{Deeply Virtual Compton Scattering}

For the DVCS analysis~\cite{eps99papers},
events with only two EM clusters and at most one track (which must be 
matched to one of the clusters) are selected. The first (second) candidate, 
corresponding to the scattered $e^+$ ($\gamma$) in the DVCS case, must have 
polar angle $\theta_1>2.8$ ($\theta_2<2.4$) radians and 
$E_1 > 10$~GeV ($E_2 > 2$~GeV). To suppress the QEDC process the polar 
angle difference must satisfy $|\theta_1-\theta_2|>0.8$ radians. Among the 
remaining requirements are a cut of $Q^2>6$~GeV$^2$, 
calculated using the first EM 
candidate, and a cut on the invariant mass of the 
two EM candidates, $M_{12}<30$~GeV. From $37{\rm pb}^{-1}$ of $e^+p$ data,
1954 events remain for further study after application of all cuts.

As an aid to studying DVCS, a MC generator GenDVCS~\cite{gendvcs} 
based on the DVCS, QEDC, and interference term (int) 
cross sections provided by FFS~\cite{ffs} was developed at ZEUS. 
Samples of DVCS+QEDC+int events (elastic only) were generated 
according to the predicted cross sections, run through a full detector 
and trigger simulation, and processed using the same reconstruction program 
as the real data. Additional samples were similarly generated using the 
QEDC generator Compton2.0~\cite{compton20} (elastic only) for comparison,
as well as RAPGAP~\cite{rapgap} (diffractive events) and  
DJANGOH~\cite{djangoh} (inclusive DIS events) samples for studying background
from $\pi^0/\eta$ contamination.

\begin{figure}
\centerline{
%\hspace{0.1cm}
\epsfxsize=7.5cm
\epsfbox{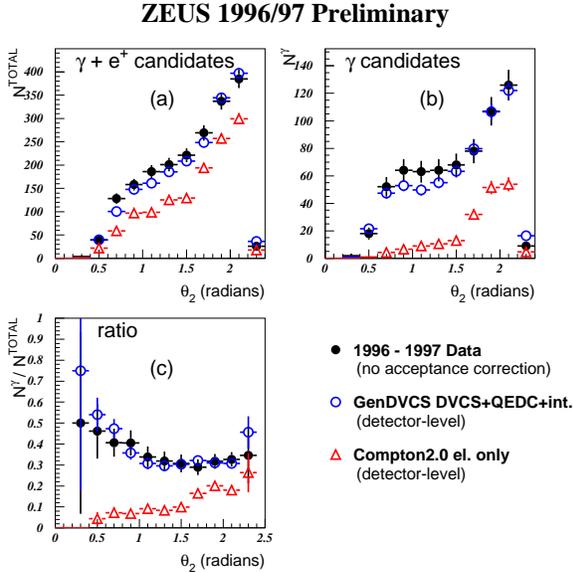}}
\caption{Distribution of $\theta_2$ for the DVCS sample.}
\label{fig_dvcs_theta2}
\end{figure}

Shown in Fig.~\ref{fig_dvcs_theta2} is the polar angle distribution of the
second EM candidate, $\theta_2$, for (a) all candidates, and (b) only 
those candidates without a matched track ($\gamma$ candidates). The third plot
(c) shows the ratio of (a) and (b). Overlaid are the predictions from 
Compton2.0 and GenDVCS, normalised to the same luminosity as the data, based 
on their calculated cross sections. 
(The QEDC ratio is unchanged when the inelastic component is included.)
There is a clear deficit in the number of small angle 
$\gamma$ candidates predicted by the QEDC simulation. The inclusion of DVCS
brings the prediction into reasonable agreement with the data. 

\begin{figure}
\centerline{
\epsfxsize=8.cm
\epsfbox{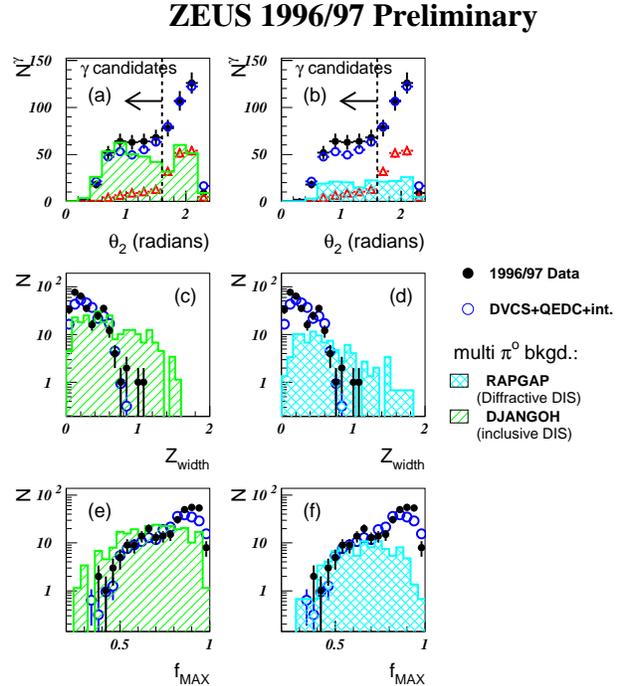}}
\caption{Shower shapes for DVCS $\gamma$ candidates.}
\label{fig_dvcs_shape}
\end{figure}

A potential source of background arises from processes like
$e^+p \rightarrow e^+\pi^0 p$, etc., where a $\pi^0$/$\eta$ fakes a 
$\gamma$ signal in the calorimeter. Redisplayed 
in Figs.~\ref{fig_dvcs_shape}(a)
and (b) is the $\theta_2$ distribution for $\gamma$ candidates.
Overlaid are the RAPGAP and DJANGOH predictions. It may thus appear possible 
to explain the data as being due to such events.
However, these two generators are not expected to 
predict accurate rates for low-multiplicity $\pi^0/\eta$ production. 
In fact, as shown in
Figs.~\ref{fig_dvcs_shape}(c)-(f), 
the $Z_{width}$ and $f_{max}$ distributions for BCAL 
$\gamma$ candidates having $\theta_2<1.6$ radians, where the QEDC 
contribution is predicted to be small, indicate that the
$\pi^0/\eta$ hypotheses cannot account for the EM shower shapes,
and so the data cannot be 
explained as hadronic background from low-multiplicity DIS events.
A repetition of the analysis~\cite{eps99papers} 
with a harder energy cut ($E_2 > 5$~GeV) on the 
second candidate further supports this conclusion. This, then, is 
first evidence for DVCS at high energy.

\section{Acknowledgements}

I gratefully acknowledge the many useful discussions about DVCS I have had 
with A.~Freund and M.~Strikman. I thank the conveners of my session for 
their kind and considerate handling of the scheduling of my talk.

\end{document}